# Trust! Why it Has Been Lost and How to Regain It

Didier Sornette

Chair of Entrepreneurial Risks
Department of Management, Technology and Economics
ETH Zurich, Zurich, Switzerland

Swiss Finance Institute
c/o University of Geneva, 40 blvd. Du Pont d'Arve
CH 1211 Geneva 4, Switzerland

Competence Center for Coping with Crises in Socio-Economic Systems
ETH Zurich, Zurich, Switzerland
(http://www.ccss.ethz.ch/)

*This essay suggests that a proper assessment of the presently unfolding financial crisis, and its cure, requires going back at least to the late 1990s, accounting for the cumulative effect of the ITC, real-estate and financial derivative bubbles. We focus on the deep loss of trust, not only in Wall Street, but more importantly in Main Street, and how to recover it on the short and long terms. A multi-disciplinary approach is needed to deal with the nonlinear complex systems of the present world, in order to develop a culture of fairness, and of upside opportunities associated with a risky world.*

JEC: D00, E00, F00, G00, H00, O00

The on-going credit crisis and panic shows that financial price and economic value are based fundamentally on trust; not on fancy mathematical formulas, not on subtle self-consistent efficient economic equilibrium; but on trust in the future, trust in economic growth, trust in the ability of debtors to face their liabilities, trust in financial institutions to play their role as multipliers of economic growth, trust that your money in a bank account can be redeemed at any time you choose. Usually, we take these facts for granted, as an athlete takes for granted that her heart will continue to pump blood and oxygen to her muscles. But what if she suffers a sudden heart attack? What if normal people happen to doubt banks? Then, the implicit processes of a working economy –all we take for granted– starts to dysfunction and spirals into a global collapse.

Many observers and pundits attribute the present mess to the mortgage-backed securities linked to the bursting of the house price bubble, the irresponsible lending, overly complex financial instruments and conflicts of interest leading to asymmetric information translated into market illiquidity, and the spreading of risks via packaging and selling of imagined valuations to unsuspecting investors.

Here, I would like to suggest that a proper assessment, and the cure, requires going back at least to the late 1990s, accounting for the interplay of three successive bubbles.

## The three bubbles

Ten years back, the ITC "new economy" bubble was expanding at full steam. The 1990s led to a roughly doubling of the GDP growth rate of the US compared to Europe, a feast attributed to comparable gain in productivity that was thought to be permanent. However, that turned out to be a bubble that crashed in the first half of 2000, triggering two dismal years for the stock market (losses larger than 60% from top to bottom from 2000 to 2003) and a (mild) recession.

To fight the recession and the negative economic effects of a collapsing stock market, the Fed engaged in a pro-active monetary policy (decrease of the Fed rate from 6.5% in 2000 to 1% in 2003 and 2004) which, along with expansive Congressional real-estate initiatives, fueled what can now be rated as one of the most extraordinary real-estate bubbles in history, with excesses on par with those that occurred during the famous real-estate bubble in Japan in the late 1980s. As Alan Greenspan himself documented in a scholarly paper researched during his tenure as the Federal Research Chairman at that time, the years from 2003 to 2006 witnessed an extraordinary acceleration of the amount of wealth extracted by Americans from their houses. The negative effects on consumption and income due to the collapse of the first ITC bubble were happily replaced by an enthusiasm and a sense of riches permeating the very structure of US society. With wanton abandon, both the public and Wall Street were the (often unconscious) actors of a third concomitant bubble inflated by subprime mortgage-backed securities (MBS) and complex packages of associated financial derivatives.

But to be clear: These financial instruments were great innovations which, in normal times, would indeed have provided a win-win situation: More people have access to loans, which become cheaper because banks can sell their risks to the supposed bottomless reservoirs of investors worldwide with varying appetites for different risk-adjusted returns.

However, the excesses culminating with the third bubble were so enormous that, as has been argued by many astute observers, too many of the MBS were "fragile": they were linked to two key unstable processes: the value of houses and the loan rates. The "castles in the air" of bubbling house prices promoted a veritable eruption of investments in MBS, these investments themselves pushing the demand for and therefore the prices of houses – until the non-sustainability of these mutually as well as self-reinforcing processes became apparent.

## Proximal cause and deep-rooted sickness

In March 2007, the first visible signs of a worrisome increase in default rates appeared, followed in the summer of 2007 by the startled realization of the financial community that the problem would be much more serious. In parallel with an acceleration of the default of homeowners and a deteriorating economic outlook, the second part of 2007 and first part of 2008 saw a rapid increase, punctuated by dramatic bankruptcies, in the estimated cumulative losses facing banks and other investment institutions: from initial guesses in the range of tens of billions, to hundreds of billions, then to more than a trillion. The Federal Reserve and Treasury stepped up their actions in proportion to the ever-increasing severity of the uncovered failures, unaware of the enormity of the underlying imbalances, or unwilling to take the measures that would address the full extent of the problem, which only now (one hopes) has revealed its enormous systemic proportions. Let us be blunt: Government has been blissfully unaware of the really predictable cumulative effect of these three bubbles, as each one appeared to "solve" the problems induced by the previous one. Monetary easing, the injection of liquidity, successive bailouts, all address symptoms and ignore the sickness.

The sickness is the cumulative excess liability present in all the sectors of the US economy (debts of U.S. households as a percentage of disposable income at around 130%, those of U.S. banks as a percentage of GDP currently around 110%, U.S. Government debt at 65% of GDP, corporate debts at 90% GDP, state and local government debts at 20% GDP, unfunded liabilities of Medicare, Medicaid, and Social Security in the range of 3-4 times GDP). Such levels of liabilities in the presence of the bubbles have produced a highly reactive unstable situation, in which sound economic valuation becomes unreliable, further destabilizing the system. This sickness has only been worsened by measures that disguise or deny it, like the misapplied innovations of the financial sector, with their flawed incentive structures, and the de facto support of an all-too willing population, eager to believe that wealth extraction could be a permanent phenomenon. On the sustainable long term, the growth of wealth has to be equal to the actual productivity growth, which is about 2-3% in real value on the long term in developed countries.

## The loss of trust

Because of this failed governance, the crisis has accelerated, now in a burst of such intensity that it has forced coordinated actions among all major governments. While their present involvement may restore short-term confidence, much more is needed to address the depth of the problem. At one level, the loss of trust between financial institutions stems from the asymmetric information on their suddenly escalating counter-party risks, making even the most solid bank perceived as a potential candidate for default, leading to credit paralysis. This can be addressed by financial measures and market forces. At a deeper level, people in the street have lost confidence, by observing a succession of problems, timidly addressed by decision makers proposing ad hoc solutions to extinguish the fire of the day (Bear Stearns, Fannie, Freddie, AIG, Washington Mutual, Wachovia), with no significant result but only more deterioration.

Nothing can resist loss of trust, since trust is the very foundation of society and economy. That people haven't yet made a run on the banks is not, given today's insurance policies against the catastrophes of the past, sufficient indication to the contrary. In fact, there may already be an "invisible run on the banks", as electronic and wire transfers have been accelerating in favor of government-backed Treasury bills. A significant additional impediment to the restoration of public trust is that the Fed, Treasury and concerted government actions are perceived as supporting that "gains are private while losses are socialized."

Present actions attempt to stabilize the financial sector by making governments, therefore taxpayers, the lenders and buyers of last resort. But collectively people are more intelligent than governments and decision makers think. They know that governments, in particular in the West, have not saved (counter-cyclically a la Keynes) during the good years, and they thus wisely doubt their prudence during the bad ones. They suspect that their governments will eventually extract the needed capital from them. They suspect intuitively that the massive measures taken to support the financial world will do little to help general economies of the US, Europe and the rest of the world.

In the short-term, these programs by governments and central banks are perhaps necessary evils, to prevent the serious consequences for the real economy if the credit problem is not solved, which include spiraling down home prices, growing unemployment, and sluggish capital spending. But if not complemented by additional pro-active measures, they miss the fundamental goal, which is to restore trust.

## How to restore trust

I envision the restoration of trust and the transition to a credible sustainable regime from the following actions. While guarantees and direct equity investments seem better tools than acquisition of bad debts at above-market prices, I only discuss here the actions which go beyond the short-term and which have been mostly absent in the debates.

First, governing bodies must for once play to the intelligence of the crowd. What needs to be done is to explain truthfully (is this possible nowadays?) the true cause of the problems: the more than one decade of excesses and the three successive and inter-related bubbles, the fact that the liabilities are enormous and that the budget has in the end to be balanced, that accelerating borrowing on the future cannot be a sustainable strategy. As humans, we are more inspired to trust when failures are acknowledged than when blame is shifted to someone else. This is the core reason why going to the fundamental source of the problems may be part of the solution in restoring confidence in the existence of a solution at minimal cost.

Second, the issue of fairness is essential for restoring confidence and support. Banks have acted incompetently in the recent bubble by accepting package risks, by violating their fiduciary duties to the stockholders, by letting the compensation/incentive schemes run out of control. There is an absolute need to rebuild that confidence, and this applies also to the regulators. This requires new strong regulations to deal with conflicts of interest, moral hazard, and to enforce the basic idea of well-functioning markets in which investors who took risks to earn profits must also bear the losses. For instance, to fight the rampant moral hazard that fueled the bubbles, share-holders should be given "clawback" permission, that is, the legal right to recover senior executive bonus and incentive pay, that proved to be ill-founded. In addition, many of the advisors and actors of the present drama have vested interest and strong conflict of interests, including the Fed, the Treasury and the major banks acting on behalf or with the approval of the Treasury. An independent elected body would be one way to address this problem, ensuring separation of interest and power.

## Regulations: necessary but "law of unintended consequences"

The three bubbles provide vivid support for the Minsky hypothesis that, by leveraging opportunities in speculative euphoria, financial markets produce endogenous instabilities. Contrary to standard economic theory that views crises as only caused by exogenous shocks which are inherently unpredictable, endogenous crises are predictable, which makes the existence of the present turmoil all the more upsetting as it could have been mitigated.

Does this mean that, when out of this crisis, regulations and interventions should be devised to control these predictable outcomes of financial market dynamics? Our study of many historical cases suggests that some bubbles are actually quite beneficial on the long-term by their creation of over-capacity in new technology, resulting from a social climate pushing the investors to take risks that they would never have taken otherwise by using a rational cost-benefit analysis. Examples abound, from the railway boom of 1840 in the UK to the recent ITC bubble bursting in 2000. In the former case, the over-capacity of the railway system gave the UK an enormous first-to-scale advantage in the burgeoning industrial revolution compared with the competing French and German powers. The ICT bubble led also to a lot of short-term destruction of virtual money (the unrealistic tripling of "new technology" stock values in 1998 and 1999) but this was accompanied with important innovations in the burgeoning Internet world, that society is now progressively learning to profit from and exploit. Bubbles may thus play a useful role, as the excess capacity built during their excesses allows novel economic regimes to emerge. Regulation should not try to fight bubbles per se, only the well-understood mechanisms and incentives that lead to artificial constructions without real long-term value, which are associated with as moral hazard, conflicts of interest and asymmetric information.

Regulations are the natural response taken by governments in times of crises, but there are many problems with them. Regulations are necessary to ensure fairness and structural stability (while short-term swings are not necessarily nefarious and should not be combated by all means, as return is the compensation for taking risks). The problem is that most regulations are either too simple or too complex to provide the intended benefits. The main difficulty with regulation lies in the "law of unintended consequences" or "illusion of control": all regulations have negative secondary effects, which are almost never foreseen. The Sarbanes-Oxley Act of 2002 enforcing enhanced standards for all U.S. public company boards, management, and public accounting firms, is a case in point: the cost of being a publicly held company in the U.S. increased by 130 percent, making many small businesses and foreign firms to deregister from US stock markets, at the benefit of foreign centers. In addition, the complexity of the new accounting reports made the information even less transparent, debasing the very goal of the Act. As another recent example, consider the mortgage rates which have soared from 5.87% to 6.38% during the week of 13 Oct. 2008 in an unexpected reaction to the latest Treasury financial rescue plan: In response to its announcement that it would take equity stakes in banks and guarantee new bank debt, investors were prompted to buy bank debt and sell bonds backed by home loans, making access to loans even more expensive for households.

There is however a domain where, in hindsight, regulations would have had great benefits in defusing the present crisis: financial derivatives. Again going back the 1990s, Alan Greenspan, supported successively by then Treasury Secretaries Robert Rubin and Laurence Summers, convinced the U.S. Congress to make the fateful decision not to pass any legislation that would have supervised the development and use of financial derivatives, notwithstanding various attempts by legislators and the call from expert financiers of the caliber of Warren Buffet and Georges Soros who warned years before the present crisis about these "weapons of financial mass destruction". After being one of the most vocal supporters of the self-regulation efficiency of financial markets, Alan Greenspan is now writing in his memoirs that the villains were the bankers whose self-interest he had once bet upon for self-regulation. This brings us back full circle to what I consider a key issue: regulations should aim at ensuring a culture of integrity and ethical behavior. The most advanced research in behavioral and experimental economics in particular informs us that harsh punishments and the promotion of social culture norms are key mechanisms for building morality and cooperation.

**A culture of risks, opportunities and robustness**
Beyond the immediate concerns, we need to keep in mind the big picture, that this time is a unique opportunity for change and improvement. The present crisis should be exploited as a unique opportunity to start developing a genuine culture of risks, which should be obligatory training for managers in governments, in regularity bodies, and in financial institutions. One little discussed reason for the present crisis is the lack of adequate education of top managers on risks in all its dimensions and implications. This is also the time that a culture of risks starts permeating the public at large. In the 21$^{st}$ century, "linear thinking" should be replaced by a growing appreciation of the inter-connectivity and feedbacks of the complex systems we deal with. Recent advances in complex system theory provide novel ways to think and act, in particular to ensure more robustness and resilience. Key principles, often in contradiction with those underlying globalization, include the need for variety, redundancy, compartments, sparseness of networks and consequences of the unavoidable delays and synchronization of actions. This "robustness" approach is well exemplified by Warren Buffet's philosophy of investing, based on "achieving acceptable long-term results under extraordinary adverse conditions," which contrasts with standard financial practices based on estimated likelihoods for meeting obligations.

What also needs to be explained is that it is normal, natural and it should be expected that systems which are creative, dynamical, innovative and which take risks, are necessarily exposed to business cycles and surprising downturns. It was a myth to believe that the "new economies" could remove all these fluctuations. What needs to be developed is a culture in which risks and downturns are understood to be the norm of nature and of society (A healthy heart beats irregularly, while death is imminent when the beating becomes regular and predictable.). This does not of course excuse the excesses due to incompetence and governance failures. Trust can be restored as it is realized that the economies of our countries are fundamentally sound, that we should not fear a mild recession, that such fear would be the *cause* of a major self-fulfilling recession, and that it is our belief that creates these very outcomes that we dread.

It is however unfair that those who are going to feel most severely the squeeze of the crisis are the middle and low-income families, as a result of the cascade of bad consequences in the economy resulting from this crisis. Innovative actions should be attempted to fix the bailout bottom-up, by helping the 40 millions indebted home owners; for instance, Thomas Peterfly, CEO of Interactive Brokers Group, proposes that the Treasury pays the first $250 of every American's primary residential mortgage each month for five years.

We have been drugged by the seductive belief that constant (or even accelerating) growth, with no plateau, no consolidating phase, is possible. From the perspective offered by history, the vagaries that we see today are the amplification resulting from an epoch dominated by the political and business elites, who are focused on short-term elections and immediate revenues, rather than on developing the foundations for the long-term sustainability. Some countries, for instance Finland and Switzerland in Europe, or Singapore in Asia, have put long-term growth as their focus and they should fare this crisis much better on the medium and long term. Economic growth can only be based on innovations and gains of productivity, which derive from better education and infrastructures. Governments and decision makers should re-focus on these key issues.

The insights summarized here can help foster a reconstruction of financial institutions, based on a deeper rethinking of how to manage our economic affairs in the light of subjective and collective perceptions. The outlook can therefore in fact be promising… if we have the patience to endure duress calmly and use this crisis as an exceptional opportunity.

**References and further reading**

**Acknowledgements:** This article benefited from many feedbacks from colleagues and collaborators, including K. Axhausen, Lars-Erik Cederman, Riley Crane, Dirk Helbing, Hans Herrmann, Yannick Malevergne, Jeffrey Satinover, Jan-Egbert Sturm, Georg von Krogh and Hilary and Ryan Woodard.

*Didier Sornette is Professor of Entrepreneurial Risks at the Swiss Federal Institute of Technology, Zurich (ETH Zurich), a member of the Swiss Finance Institute, and cofounder of the Competence Center for Coping with Crises in Socio-Economic Systems at ETH Zurich.*